\documentclass[useAMS,usenatbib,usegraphicx]{mn2e}

\def\lya{L\lowercase{y}${\alpha}$ }

\def\cgs{$\rm erg \, s^{-1} \, cm^{-2}$ }
\def\zn{$z\sim9$ }

\title[ZEN2: narrow $J$-band observations directed toward three lensing clusters]{ZEN2: A narrow $J$-band search for \zn \lya emitting galaxies directed towards three lensing clusters.} \author[J. P. Willis, F. Courbin, J.-P. Kneib, D. Minniti]{J. P. Willis$^{1}$\thanks{E-mail: jwillis@uvic.ca}, F. Courbin$^{2}$, J.-P. Kneib$^{3}$ and D. Minniti$^{4}$\\ $^{1}$Department of Physics and Astronomy, University of Victoria, Elliot Building, 3800 Finnerty Road, Victoria, BC, V8P 1A1, Canada.\\ $^{2}$Laboratoire d'Astrophysique, Ecole Polytechnique F\'ed\'erale de Lausanne (EPFL), 1290 Sauverny, Switzerland.\\ $^{3}$Laboratoire d'Astrophysique de Marseille, Traverse du Siphon BP8, 13376 Marseille Cedex 12, France\\ $^{4}$Department of Astronomy, P. Universidad Cat{\'o}lica, Av. Vicu{\~n}a Mackenna 4860, Casilla 306, Santiago 22, Chile.}
\begin{document}

\date{Accepted 2007 September 03. Received 2007 August 24; in original form 2007 July 30.}

\pagerange{\pageref{firstpage}--\pageref{lastpage}} \pubyear{2006}

\maketitle

\label{firstpage}

\begin{abstract}
We present the results of a continuing survey to detect \lya emitting
galaxies at redshifts \zn: the ZEN (``z equals nine'') survey. We have
obtained deep VLT/ISAAC observations in the narrow $J$--band filter
NB119 directed towards three massive lensing clusters: Abell clusters
1689, 1835, and 114. The foreground clusters provide a magnified view
of the distant universe and permit a sensitive test for the presence
of very high-redshift galaxies.  We search for \zn \lya emitting
galaxies displaying a significant narrow-band excess relative to
accompanying $J$-band observations that remain undetected in HST/ACS
optical images of each field.  No sources consistent with this
criterion are detected above the unlensed 90\% point-source flux limit
of the narrow-band image, $F_{\rm NB}=3.7 \times 10^{-18}$ \cgs.  To
date, the total coverage of the ZEN survey has sampled a volume at \zn
of approximately 1700 co-moving Mpc$^3$ to a \lya emission luminosity
of $10^{43} \, \rm erg \, s^{-1}$. We conclude by considering the
prospects for detecting \zn \lya emitting galaxies in light of both
observed galaxy properties at $z<7$ and simulated populations at
$z>7$.
\end{abstract}


\section{The search for the most distant galaxies}

Observations of distant galaxies provide a direct view of the early
stages of galaxy evolution as well as probing the physical conditions
of the high-redshift intergalactic medium (IGM). The advent of the
Advanced Camera for Surveys (ACS) deployed on the Hubble Space
Telescope (HST) in addition to wide field optical cameras operated on
8m class telescopes has provided access to relatively large samples of
galaxies located at $z \sim 6$ and beyond. Observations of distant
Lyman drop-out galaxies (e.g. \citealt{sbm03}; \citealt{dick04}) and
\lya emitting galaxies selected via narrow-band photometry
(\citealt{rhoads03}; \citealt{hu04}) indicate that these bright,
distant galaxies cannot be the sole agents of the global re-ionisation
demonstrated by studies of Gunn-Peterson absorption in high-redshift
quasars \citep{bunker06}. Current solutions to this dilemma centre
upon the potential contribution of faint galaxies at $z\sim6$ \---\
pointed to in the very deepest, yet numerically smallest, ACS samples
\---\ or the possibility that an earlier epoch of more intense star
formation was responsible for the observed re-ionisation.

Perhaps the most important contribution to the debate concerning the
physical nature of bright, high-redshift galaxies has come from
Spitzer space telescope observations of rest-frame optical emission in
these systems (e.g. \citealt{eyles05}; \citealt{yan06}). Optical to
near infrared (NIR) observations of $z \sim 6$ galaxies sample
rest-frame emission blueward of the 4000\AA\ discontinuity and are
thus sensitive primarily to the spectral contribution from younger
stellar populations. The addition of Spitzer/IRAC bands, particularly
at 3.6\micron\ and 4.5\micron, samples the potential contribution of
older stellar populations in these galaxies. From the limited number
of bright $z \sim 6$ galaxies studied to date, there has emerged a
picture of their stellar populations as being relatively old (up to
700~Myr) and massive (up to $3 \times 10^{10}\, \rm M_\odot$). The
extrapolation of these integrated star formation histories to earlier
cosmic times points to an epoch of potentially intense star formation
in the predecessors of bright $z \sim6$ galaxies extending to
redshifts $z\sim 10$.

Though compelling evidence points to the existence of actively star
forming galaxies at $z>7$, the direct observation of such sources is
far from straightforward \---\ mainly due to the extreme faintness of
high-redshift galaxies. The brightest galaxies observed at $z\sim6$
display total AB magnitudes of order 24 (e.g. \citealt{sbm03}; Hu et
al. 2004).  Galaxies at $z>7$ \---\ including current samples of
candidate systems \---\ can be reasonably expected to display signal
levels $\rm AB > 25$ in NIR wavebands \citep{bouwens04}.  Obtaining a
spectroscopic redshift for such faint, continuum selected systems with
currently available technology is challenging (though not impossible,
c.f. \citealt{kneib04}).  Studies employing narrow-band filters or
long-slit spectral observations are sensitive to the subset of
high-redshift galaxies that emit a significant fraction of their
energy in the form of narrow spectral lines \---\ typically the \lya
emission line.  Deep narrow-band imaging and subsequent spectroscopic
observations have been employed successfully to generate high spectral
completeness samples of distant galaxies at redshifts $z=5.7$
\citep{hu04} and $z=6.5$ \citep{taniguchi05}. Long-slit spectral
observations of massive lensing clusters have been employed to
investigate the faint end of the \lya emission luminosity function
over the interval $4.5<z<5.7$ \citep{santos04a} in addition to
generating samples of candidate systems in the interval $7<z<10$
\citep{stark07}. In comparison to narrow-band observations, long slit
spectroscopic observations of \lya emitting sources typically probe a
lower background per resolution element yet are normally restricted to
relatively small volume studies due to the limited field of view of a
spectrograph slit.

A critical unknown factor determining the visibility of the \lya
emission feature in $z>7$ galaxies is the physical state of the
intervening IGM. Absorption studies of $z\sim6$ quasars appear to have
identified the very end of the global re-ionisation process
\citep{fan06} while measurement of the optical depth of electron
scattering at large angular scales in the CMB points to a typical
(though not definitive) epoch of reionisation around $z=11$
\citep{page06}. Clearly then, galaxies at redshifts $7<z<11$ may be
located in a partly ionised IGM where the local fraction of neutral
hydrogen around individual galaxies may be sufficient to attenuate the
\lya emission signature.  However, individual \lya emitting galaxies
have been identified in deep, narrow $z$-band surveys and confirmed
spectroscopically at redshifts $z=6.56$ \citep{hu02} and $z=6.96$
\citep{iye06}. Prompted by such observations, numerous theoretical
studies have been undertaken to compute the escape fraction of
ultra-violet (UV) photons from a volume of high-redshift intergalactic
hydrogen ionised as a result of star formation occuring in an
embedded galaxy (\citealt{haiman02}; \citealt{santos04b};
\citealt{barton04}), While the detailed properties of the \lya line
transmitted through such a medium are necessarily model dependent,
e.g. depending upon the mass, metallicity, star formation rate and
initial mass function of the burst and the local density of the IGM, a
range of plausible scenarios exist whereby a H{\sevensize II} region
of sufficient size is created such that transmission of a partially
attenuated \lya line occurs.

The above considerations serve as the motivation for a NIR search for
\lya emitting galaxies at $z>7$.  The search technique employs a
narrow $J$--band filter centred on 1.187\micron\ and is sensitive to
the signature of \lya emitting galaxies located at a redshift $z=8.8$
(termed \zn in the following text).  The remaining sections are
organised as follows: in Section \ref{sec_zen} we describe in further
detail the construction of the narrow-band survey. In Section
\ref{sec_data} we describe the techniques used to process the data and
identify candidate \zn \lya emitting galaxies. Finally in Section
\ref{sec_res} we determine the co-moving volume at \zn sampled in
terms of the \lya emission luminosity and compare this to a reasonable
range of expected properties of \zn galaxies. Throughout this paper,
values of $\Omega_{\rm M,0}=0.3$, $\Omega_{\Lambda,0}=0.7$ and $H_0=70
\rm \, kms^{-1} \, Mpc^{-1}$ are adopted for the present epoch
cosmological parameters describing the evolution of a model
Friedmann-Robertson-Walker universe. All magnitude information is
quoted using AB zero point values.

\section{The ZEN survey}
\label{sec_zen}

The desirability of detecting very high redshift \lya emitting
galaxies forms the motivation for the ``$z$ equals nine'' (ZEN)
survey. \lya emitting galaxies occupying narrow redshift intervals at
$z>7$ will present a characteristic emission excess signature in
infrared photometry employing a combination of narrow and broad band
filters. Narrow-band, NIR filters tuned to sample regions of night sky
emission devoid of strong terrestrial line features provide access
to relatively low background signals and thus permit sensitive imaging
observations to be executed. In what we refer to as ZEN1 we
constructed a 32 hour on-sky image of the Hubble Deep Field South
(HDF-S) in the narrow-band filter NB119 employing the VLT/ISAAC
facility (\citealt{paper1}; WC05 hereafter). Using deep, archival
images of the field consisting of VLT/ISAAC $J_s$-band and HST/WFPC2
optical bands we were able to execute a sensitive test for faint
narrow-band excess sources (i.e. $\rm NB < 25.2$, $J_s-\rm NB >0.3$)
that remain undetected in optical bands \---\ a practical definition
for candidate \zn \lya emitters. No candidate ZEN sources were
identified in these observations. However, the study demonstrated that
interloping low redshift emission excess sources could be successfully
identified and rejected using deep optical images and that a detailed
\zn volume selection function could be computed in terms of \lya
emission luminosity.

As part of a new study \---\ dubbed ZEN2 \---\ we have obtained
further narrow and broad band images directed toward the fields of
three low redshift galaxy clusters: A1689, A1835 and AC114 (Table
\ref{tab_fields}).  Each cluster acts as a gravitational lens and
provides a spatially magnified view of the background universe.  When
considering unresolved sources the effect of this magnification is to
increase the total brightness measured within a photometric detection
aperture.  We therefore use the presence of each cluster along the
line of sight to provide a gravitational ``boost'' to the signal from
putative \zn galaxies.  The properties of each cluster have been
described in detail in the literature and each possesses a
well-determined gravitational lens model.  Lens models describing the
clusters A1689, A1835 and AC114 are presented respectively in Limousin
et al. (2007), Smith et al. (2005) and Campusano et al. (2001).
Infrared observations of the three clusters were obtained with the
VLT/ISAAC facility as part of ESO programmes 070.A-0643, 071.A-0428,
073.A-0475 and are summarised in Table \ref{tab_fields}. Optical
observations of the three clusters consist of HST/ACS F850LP mosaics.
These are described briefly in Table \ref{tab_hst} and in further
detail in \citet{broad05} for A1689 and \citet{hempel07} for A1835 and
AC114.  The NIR observations described in this paper all fall
completely within the ACS mosaic areas.
\begin{table*}
\begin{minipage}{180mm}
\caption{NIR data obtained for the three cluster fields.}
\label{tab_fields}
\begin{tabular}{lccccccccccc}
\hline
Cluster & $\alpha$ (J2000) & $\delta$ (J2000) & Redshift & \multicolumn{4}{c}{NB119 observations} & \multicolumn{4}{c}{$J$--band observations} \\
&&&& $N_{exp}$ & DIT (s) & NDIT & $t_{exp}$ (s) & $N_{exp}$ & DIT (s) & NDIT & $t_{exp}$ (s) \\ \hline \hline
A1689 & 13:11:30.1 & $-01$:20:17.0 & 0.18 & 54 & 110 & 3 & 17820 & 57 & 35 & 4 & 7980 \\
A1835 & 14:01:02.0 & $+02$:51:46.7 & 0.25 & 82 & 100 & 3 & 24600 & 40 & 45 & 3 & 5400 \\
AC114 & 22:58:47.7 & $-34$:48:04.1 & 0.06 & 78 & 100 & 3 & 23400 & 40 & 45 & 3 & 5400 \\
\hline
\end{tabular}
\end{minipage}
\end{table*}
\begin{table}
\caption{Properties of the HST/ACS F850LP images of each field.}
\label{tab_hst}
\begin{tabular}{lccc}
\hline
Cluster & $t_{exp}$ (s) & Image scale & $5\sigma$ limiting magnitude \\
&&(\arcsec/pix$^{-1}$) & within 0\farcs7 aperture \\
\hline \hline
A1689 & 28600 & 0.05 & 27.48 \\
A1835 & 18220 & 0.04 & 26.92 \\
AC114 & 18368 & 0.05 & 26.98 \\
\hline
\end{tabular}
\end{table}

\section{Data reduction and analysis}
\label{sec_data}

The NIR NB119 and $J$--band data were processed using techniques
essentially identical to those described in WC05 and we summarize them
briefly here. Imaging data were a) dark subtracted using standard
calibration frames, b) corrected for varying pixel response using
twilight sky exposures, c) sky-subtracted having masked array regions
containing objects above a specified ADU level, d) corrected for both
high- and low-frequency spatial artefacts, and e) shifted to a common
pixel scale and coadded using a suitable pixel weighting and rejection
algorithm. The image quality in each field and filter combination was
computed as the mean full-width at half maximum (FHWM) of a sample of
bright, stellar sources visible in each image and is displayed in
Table \ref{tab_iq}.

\begin{table*}
\begin{minipage}{180mm}
\caption{Image quality and photometric properties of each reduced image.}
\label{tab_iq}
\begin{tabular}{lcccccccc}
\hline
Cluster & \multicolumn{2}{c}{Image quality (\arcsec)} & \multicolumn{2}{c}{AB Depth at 90\%\ completeness} & \multicolumn{2}{c}{Correction to total magnitudes} & \multicolumn{2}{c}{\lya flux sensitivity} \\
& && \multicolumn{2}{c}{(0\farcs7 aperture)}&&& \multicolumn{2}{c}{($\times 10^{-18}$ ergs s$^{-1}$ cm$^{-2}$)}\\

& NB119 & $J$-band & NB119 & $J$-band & NB119 & $J$-band & image plane & source plane \\ \hline \hline
A1689 & 0.47 & 0.49 & 24.5 & 24.9 & 0.63 & 0.68 & 3.7 & 0.48\\
A1835 & 0.42 & 0.42 & 24.5 & 24.8 & 0.56 & 0.55 & 3.6 & 2.38\\
AC114 & 0.45 & 0.42 & 24.5 & 24.8 & 0.63 & 0.60 & 3.8 & 2.67\\
\hline
\end{tabular}
\end{minipage}
\end{table*}

The $J$-band observations were obtained under photometric conditions
and were placed on an absolute flux scale using observations of
standard stars taken with the science data. Reference magnitudes for
standard stars observed with the NB119 filter are not available. The
NB119 data were therefore placed on an absolute flux scale by
comparing the flux measured within a 5\arcsec\ diameter circular
aperture for a sample of bright, isolated stellar sources visible in
both the $J$-band and NB119 images of each field. The colour term
$J-\rm NB$ for these reference sources is approximately zero and the
relationship between $J$-band reference magnitudes and NB119
instrumental magnitudes is linear and displays a gradient of unity. A
total of 12, 16 and 16 such reference sources were employed for the
fields A1689, A1835 and AC114 respectively.

Source detection and photometry were performed on each image using the
{\tt SExtractor} software package (Bertin and Arnouts 1996).  Once
again, we continue the approach outlined in WC05, i.e.  source
detection is optimised for the detection of marginally resolved or
unresolved objects and source fluxes are measured in circular
apertures of diameter 0\farcs7. The required correction to convert
aperture photometry to pseudo-total photometry (assumed to be a
5\arcsec\ diameter circular aperture) was computed for bright stellar
sources in each field. Corrections generated for each image are
given in Table \ref{tab_iq}, though it should be noted that
subsequent calculations based upon image photometry all use the
0\farcs7 aperture values.

\begin{figure}
\includegraphics[width=65mm,angle=270]{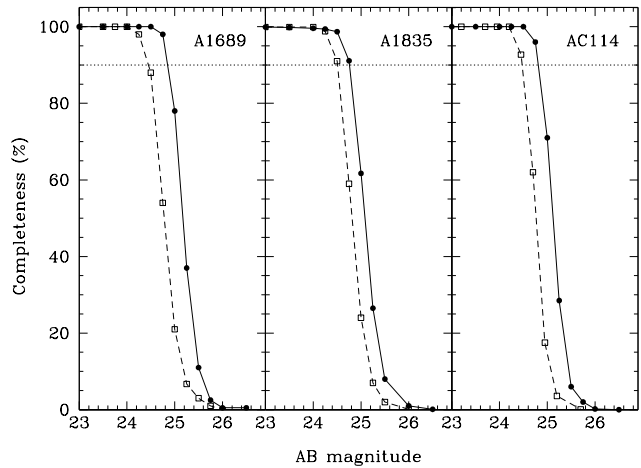}
\caption{Mean detection probability of simulated point sources as a
function of AB magnitude within the NB119 (square symbols and dashed
line) and $J$ images (circular symbols and solid line). The
horizontal dotted line indicates the 90\%\ completeness threshold.}
\label{fig_complete3}
\end{figure}

The magnitude corresponding to the 90\% point source recovery
threshold in each image ($m_{90}$) was computed by introducing
artificial point sources into the reduced image of each field and
determining the fraction recovered (see WC05). This procedure produces
a low resolution (3\arcsec\ pixels) image of the varying depth of each
image. In addition, this process generates a map of photometric
uncertainty across each field taking into account contamination by
bright galaxies. The average value of the completeness across each
image is also displayed in Figure \ref{fig_complete3} as a function of
source magnitude. The magnitudes corresponding to the 90\%\
completeness limit in each field and filter combination are displayed
in Table \ref{tab_iq}. The typical signal-to-noise ratio (SNR) of a
source displaying $m_{90}$ in each field is approximately 15. Note
that for the purposes of computing the survey selection function in
Section \ref{sec_res} we employ the two dimensional completeness
information available for each field.

\begin{figure*}
\includegraphics[width=130mm,angle=270]{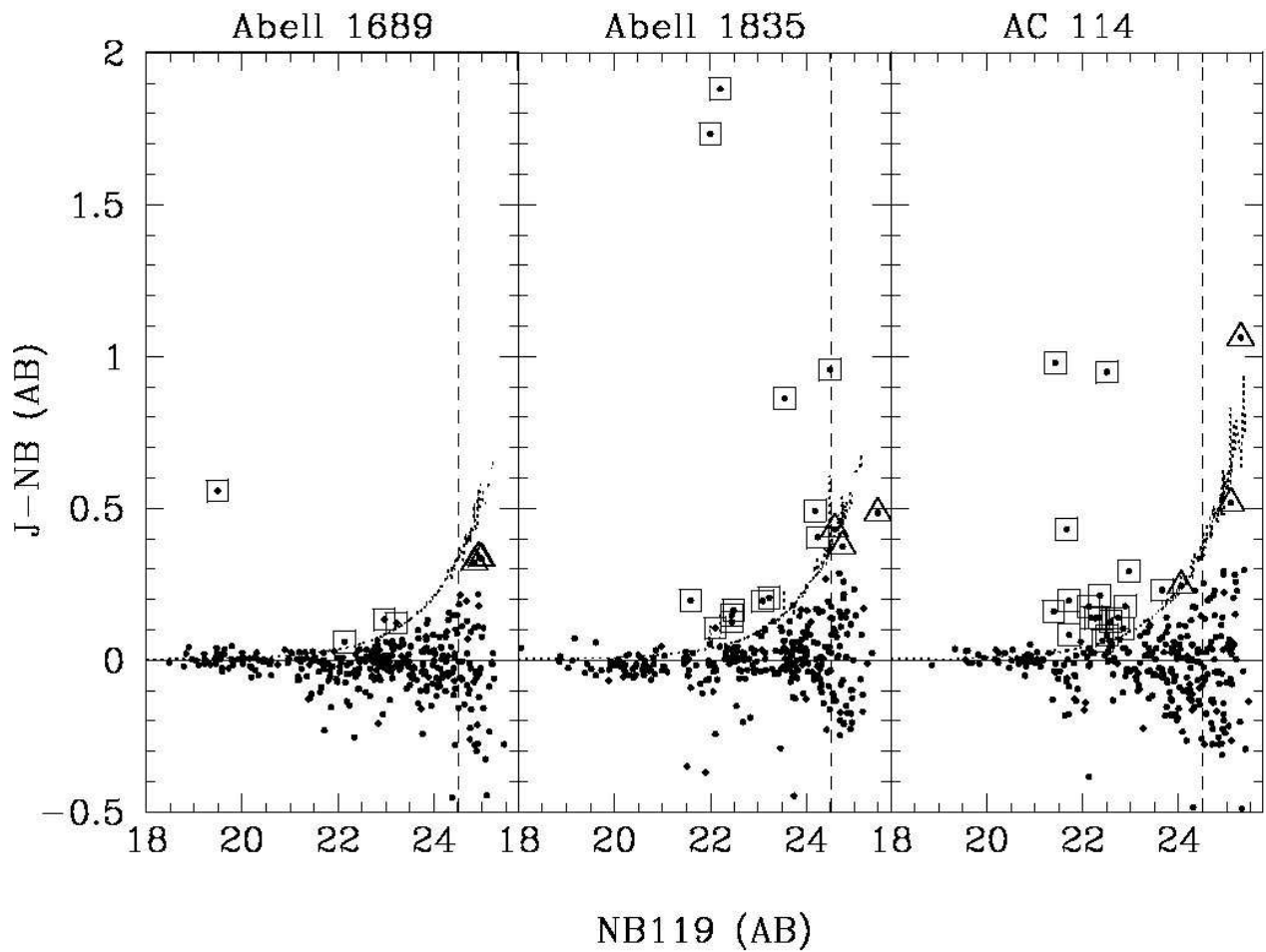}
\caption{Distribution of narrow-band excess $J-\rm NB$ versus NB119
magnitude for the three cluster fields.  Values of $J-\rm NB$ versus
NB119 for sources in each field are indicated by dots.  For each panel
the dotted curve indicates the three sigma uncertainty in the narrow
band excess and the solid horizontal line indicates $J- \rm NB=0$.The
vertical dashed line in the each panel indicates the value
NB119$_{90}$.Sources displaying a narrow-band excess greater than the
three sigma uncertainty are highlighted using a box. For clarity only
sources satisfying $\rm NB>22$ or notable brighter sources are marked
in this manner. Sources failing the above narrow-band excess measure,
yet close to either the selection envelope or the magnitude cut, are
indicated with a triangle. These sources were also inspected visually
and excluded as potential ZEN sources due to their detection in the
corresponding ACS $z$--band image.}
\label{fig_cmd3}
\end{figure*}

Candidate \zn \lya emitting galaxies are identified as narrow-band
excess sources relative to the $J$-band reference filter. Figure
\ref{fig_cmd3} displays the narrow-band excess versus NB119 magnitude
for each of the target fields. We identify emission line sources as
those displaying a positive $J-\rm NB$ signature in excess of the
local $3\sigma$ uncertainty in $J-\rm NB$. At the 90\%\ completeness
limit of each field this corresponds to a colour excess $J-\rm NB \ge
0.3$.  Of the sources satisfying this narrow-band excess threshold,
all are ultimately detected in HST F850LP images of each field.  We
therefore associate these sources with intervening emission line
galaxies (e.g. [OII]3727, H$\beta$ or H$\alpha$) located at redshifts
that place the emission feature in the NB119 filter (c.f. WC05).  The
field of view of the NIR images of each cluster cover an area of 4
square arcminutes in each case.  Within this area the images of
clusters A1689, A1835 and AC114 respectively contain 4, 13 and 21
interloping emission line sources down to the observed frame magnitude
limit $m_{90}$ appropriate for each field.  In addition, a small
number of sources in each field are detected in the narrow-band but
remain undetected in the accompanying $J$-band.  All of these sources
were investigated and ultimately associated with faint sources in
optical HST observations. Therefore, no candidate \zn \lya emitting
galaxies have been detected in any of the three fields studied.

\section{Probing the luminosity function of \zn LAE galaxies}
\label{sec_res}

The non-detection of candidate \zn \lya emitting galaxies in the three
cluster fields may be understood by computing the total volume sampled
at each \lya luminosity.  To achieve this the survey flux sensitivity
as a function of solid angle must be corrected for the magnification
introduced by the cluster lensing potential in each field, distance
dimming and the effect of partial transmission of the \lya line by the
NB119 filter.

\begin{figure*}
\includegraphics[width=180mm]{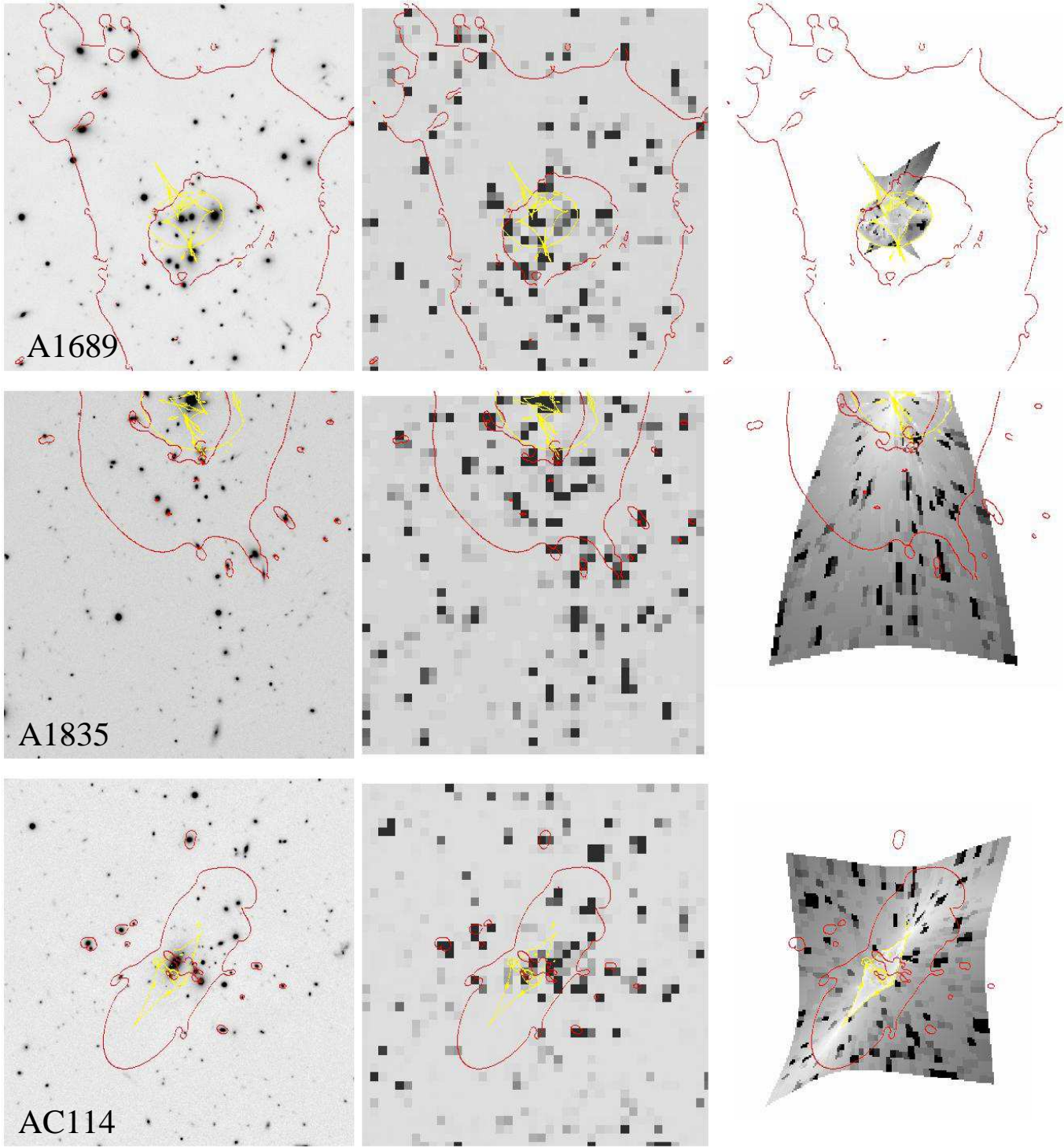}
\caption{Greyscale images demonstrating the processing applied to the
two dimensional sensitivity maps. All images measure two arcminutes on
a side with North up and East left. Panels for each cluster are
arranged in rows. Left panel: narrow-band image of each field. Centre
panel: image plane source detection sensitivity, $m_{90}$ (see text
for more details). Lighter regions indicate fainter sensitivity
levels. Right panel: \lya flux detection sensitivity computed for a
source plane located at $z=8.8$. Lighter regions indicate fainter
sensitivity levels. The field distortion arises from the inversion
of the image plane sensitivity map though the cluster potential. In
each panel the red and yellow contours indicate respectively the
critical and caustic lines corresponding to a source located at
$z=8.8$.}
\label{fig_3x3}
\end{figure*}

The gravitational potential associated with each galaxy cluster
creates a magnified view of the background universe. This
magnification can be considered as a spatially varying transformation
between the source plane and image plane geometry of a particular
field. The detection limits displayed in Figure \ref{fig_complete3}
are associated with the image plane of each cluster. The corresponding
source plane detection map for each cluster was computed using the
{\tt LENSTOOL} package \citep{kneib93} and a model potential
describing each cluster. Each pixel in the detection map (location and
area) was transformed to the source plane accounting for the varying
lens deflection angle as a function of sky position before being
reassembled onto a uniform pixel grid. Images displaying the
individual stages in this procedure are displayed in Figure
\ref{fig_3x3}.

\begin{figure}
\includegraphics[width=80mm]{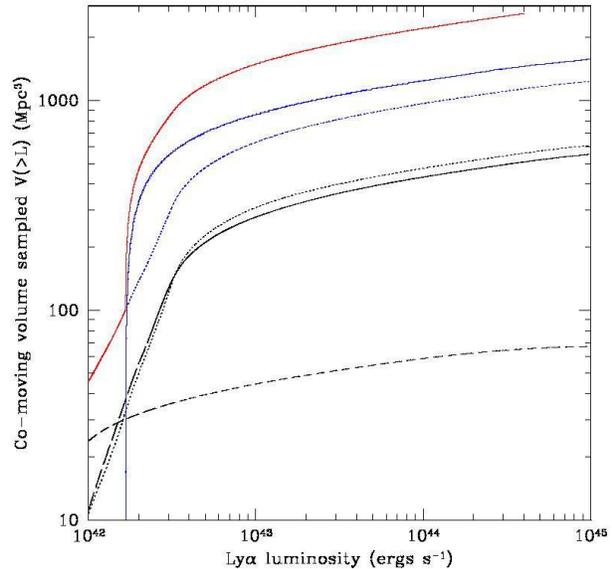}
\caption{Co-moving volume sampled brighter than a given \lya
luminosity. The black curves indicate the co-moving volume sampled
toward each of the three cluster fields in this paper: A1689 (dashed),
A1835 (dotted) and AC114 (solid). The blue curves indicate the total
co-moving volume sampled toward the three ZEN2 cluster fields (dashed)
and the co-moving volume sampled toward the ZEN1 field (HDF-South)
described in WC05 (solid). The red curve indicates the total co-moving
volume sampled by the ZEN1 and ZEN2 surveys to date.}
\label{fig_vsamp}
\end{figure}

In Appendix A of WC05 we described how the magnitude limit of a
particular narrow-band image could be transformed into a \lya emission
line luminosity at \zn by accounting for the equivalent width
criterion applied to select narrow-band excess sources and the partial
transmission of the \lya line by the narrow-band filter. The mean
\lya flux sensitivity toward each cluster field is given in Table
\ref{tab_iq} and is computed over both image plane and source plane
(i.e. de-lensed) pixels.  The volume sampled as a function of \lya
luminosity, $V(\rm L_{Ly\alpha})$, is then computed as the integral
over the differential co-moving volume element out to the maximum
redshift at which a source displaying the specified luminosity would
be detected. We apply the same procedure to determine $V(\rm
L_{Ly\alpha})$ in the current study with only a minor modification:
rather than compute the volume sampling based upon the average depth
over each field, we compute the volume sampled per pixel in the source
plane detection map describing each cluster. The volume sampled as a
function of \lya luminosity for each cluster field is then computed as
the contribution from individual detection map pixels, weighted by the
solid angle of each pixel. The volume selection function for each
cluster is displayed in Figure \ref{fig_vsamp}. Each curve is computed
assuming a \lya emission line of rest frame velocity width $\sigma_v =
50~\rm kms^{-1}$ (see WC05 for additional details).

\begin{figure}
\includegraphics[width=80mm]{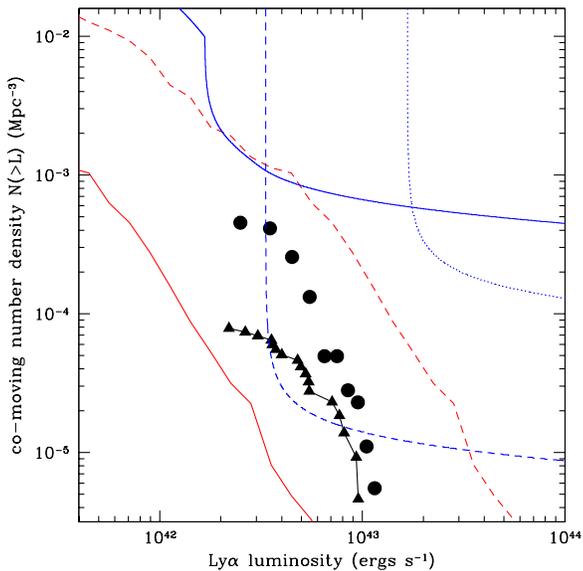}
\caption{A comparison of the limits imposed upon the $z>7$ \lya
luminosity function by existing and planned NIR narrow-band
surveys. The blue curves indicate the region of the cumulative space
density versus \lya luminosity sampled by observations: the ZEN1$+$2
surveys (solid), the wide area, shallow depth ISAAC survey of Cuby et
al. (2006) (dotted) and the planned sensitivity of ZEN3, a wide area
NB survey employing CFHT WIRCam (dashed). The points indicate the
observations of $z=6.6$ \lya emitting galaxies by \citet{kashikawa06}:
circles indicate the photometric sample corrected for completeness
while the triangles indicate the spectroscopically confirmed
sample. The red curves indicate the \lya luminosity function simulated
at $z=9$ by \citet{delliou06}, assuming a \lya escape fraction of 0.02
(solid) and 0.2 (dashed) respectively.}
\label{fig_lfcomp}
\end{figure}

The inverse of the volume selection function for the ZEN survey is
equal to the cumulative space density of \zn \lya emitting galaxies
sampled as a function of their emission luminosity (Figure
\ref{fig_lfcomp}).  The region above each curve in Figure
\ref{fig_lfcomp} indicates the region of the cumulative luminosity
function of putative \zn \lya emitting galaxies that can be ruled out
as a result of the non-detection of bona-fide \zn sources It is
instructive to compare these limits to the cumulative luminosity
function both of observed \lya emitting galaxies at redshift $z=6.6$
(Kashikawa et al. 2006) and to \zn \lya emitting galaxies simulated
within a semi-analytic model of galaxy formation (Le Delliou et al
2006). We consider the implications for each population in turn. If we
use the observed population of \lya emitting galaxies at $z=6.6$ as a
model for emission at \zn then the current areal coverage of the ZEN
survey would have to be increased by a factor of at least three in
order to provide a realistic constraint on the putative \zn
population. The prospect for extending deep, narrow-band surveys at
\zn to wider areal coverage is promising given the advent of both the
DAZLE \citep{horton04} and HAWK-I \citep{casali06} NIR cameras at the
ESO VLT. If instead the properties of \lya emitting galaxies at \zn
are described by the semi-analytic model of Le Delliou et al. then the
prospects for their detection is less certain.  A small region of the
model described by a UV photon escape fraction of 0.2 has already been
tentatively ruled out by the ZEN survey.  However, reducing the escape
fraction to 0.02 results in a proportionate decrease in the \lya
luminosity and the detection of such a population with either DAZLE or
HAWK-I will remain challenging. An interesting alternative approach
employs the current generation of relatively wide field NIR cameras
operating on 4m class telescopes. In Figure \ref{fig_lfcomp} we
display the anticipated results of ZEN3, a narrow-band search for \lya
emitting galaxies at $z \sim 8$ currently underway using the Canada
France Hawaii Telescope (CFHT) WIRCam facility \citep{puget04}. The
exceptional volume sampling of such wide field cameras will permit a
very sensitive test of the space density of putative $z \sim 8$
emitters, whether based upon observed $z=6.6$ or model populations.

%We therefore conclude that the unambiguous identification of $z>7$
%will require a sustained observational effort: employing multiple
%facilities to overcome the sampling restrictions of area versus depth
%and multiple techniques to overcome the incomplete view of the
%high-redshift galaxy population provided by a single selection
%strategy.

\section*{Acknowledgments}

The authors wish to thank both Nobunari Kashikawa and Cedric Lacey for
making their data available in electronic form. JPW acknoweldges
financial support from the Canadian National Science and Engineering
Research Council (NSERC). FC acknowledges financial support from the
Swiss National Science Fundation (SNSF). DM is partially supported by
FONDAP 15010003.

\bsp

\label{lastpage}


\begin{thebibliography}{99}

\bibitem[\protect\citeauthoryear{Barton et al.}{2004}]{barton04}
Barton, E.~J., Dav{\'e}, R., Smith, J.-D.~T., Papovich, C., Hernquist,
L., Springel, V., 2004 ApJL, 604, 1

%\bibitem[\protect\citeauthoryear{Becker et al.}{2001}]{becker01}
%Becker, R. et al., 2001, AJ, 122, 2850

\bibitem[\protect\citeauthoryear{Bertin and Arnouts}{1996}]{bertin96}
Bertin, E., Arnouts, S., 1996, A\&AS, 117, 393

\bibitem[\protect\citeauthoryear{Bouwens et al.}{2004}]{bouwens04}
Bouwens, R.~J., et al.\ 2004, ApJ, 616, L79

\bibitem[\protect\citeauthoryear{Broadhurst et al.}{2005}]{broad05}
Broadhurst, T., et al., 2005, ApJ, 621, 53

\bibitem[\protect\citeauthoryear{Bunker et al.}{2006}]{bunker06}
Bunker, A., Stanway, E., Ellis, R., McMahon, R., Eyles, L., Lacy,
M., 2006, New Astronomy Review, 50, 94

\bibitem[\protect\citeauthoryear{Campusano et al.}{2001}]{camp01}
Campusano L.,E., Pell{\`o}, R., Kneib, J.-P., Le Borgne, J.-F., Fort,
B., Ellis, R., Mellier, Y., Smail, I., 2001, A\&A, 378, 394.

\bibitem[\protect\citeauthoryear{Casali et al.}{2006}]{casali06}
Casali, M., et al.\ 2006, Proc. SPIE, 6269, 29

\bibitem[\protect\citeauthoryear{Dickinson et al.}{2004}]{dick04}
Dickinson, M., Stern, D., Giavalisco, M., 2004, ApJ, 600, 99

\bibitem[\protect\citeauthoryear{Eyles et al.}{2005}]{eyles05} Eyles,
L.~P., Bunker, A.~J., Stanway, E.~R., Lacy, M., Ellis, R.~S., \&
Doherty, M.\ 2005, MNRAS, 364, 443

%\bibitem[\protect\citeauthoryear{Fan et al.}{2002}]{fan02} Fan, X.,
%Narayanan, V.K., Strauss, M.A., White, R.L., Becker, R.H., Pentericci,
%L., Rix, H.-W., 2002, AJ, 123, 1247

\bibitem[\protect\citeauthoryear{Fan et al.}{2006}]{fan06} Fan, X. et
al., 2006, AJ, 132, 117

\bibitem[\protect\citeauthoryear{Gunn and Peterson}{1965}]{gunn65}
Gunn, J.E., Peterson, B.A., 1965, ApJ, 142, 1633

\bibitem[\protect\citeauthoryear{Haiman}{2002}]{haiman02} Haiman, Z.,
2003, ApJ, 578, 702

\bibitem[\protect\citeauthoryear{Hempel et al.}{2007}]{hempel07}
Hempel, A., et al., 2007 A\&A submitted.

\bibitem[\protect\citeauthoryear{Horton et al.}{2004}]{horton04}
Horton, A., Parry, I., Bland-Hawthorn, J., Cianci, S., King, D.,
McMahon, R., \&\ Medlen, S.\ 2004, proc. SPIE, 5492, 1022

\bibitem[\protect\citeauthoryear{Hu et al.}{2002}]{hu02} Hu, E.M.,
Cowie, L.L., McMahon, R.G., Capak, P., Iwamuro, F., Kneib, J.-P.,
Maihara, T., Motohara, K., 2002a, ApJL, 568, 75

\bibitem[\protect\citeauthoryear{Hu et al.}{2004}]{hu04} Hu, E.~M.,
Cowie, L.~L., Capak, P., McMahon, R.~G., Hayashino, T., Komiyama, Y.,
2004, AJ, 127, 563

\bibitem[\protect\citeauthoryear{Iye et al.}{2006}]{iye06} Iye, M., et
al.\ 2006, Nature, 443, 186

\bibitem[\protect\citeauthoryear{Kashikawa et al.}{2006}]{kashikawa06}
Kashikawa, N., et al.\ 2006, ApJ, 648, 7

\bibitem[\protect\citeauthoryear{Kneib}{1993}]{kneib93} Kneib,
J.-P., 1993, Ph.D. Thesis.

\bibitem[\protect\citeauthoryear{Kneib et al.}{2004}]{kneib04} Kneib,
J.-P., Ellis, R. S., Santos, M. R., Richard, J., 2004, ApJ, 607, 697

\bibitem[\protect\citeauthoryear{Limousin et al.}{2007}]{limo07}
Limousin, M., Richard, J., Jullo, E., 2007, ApJ in press
(astro-ph/0612165)

\bibitem[\protect\citeauthoryear{Le Delliou et al.}{2006}]{delliou06}
Le Delliou, M., Lacey, C.~G., Baugh, C.~M., \& Morris, S.~L.\ 2006,
MNRAS, 365, 712

\bibitem[\protect\citeauthoryear{Page et al.}{2006}]{page06} Page,
L., et al., 2006, ApJ submitted (astro-ph/0603450)

\bibitem[\protect\citeauthoryear{Puget et al.}{2004}]{puget04} Puget,
P., et al.\ 2004, Proc. SPIE, 5492, 978

\bibitem[\protect\citeauthoryear{Rhoads et al.}{2003}]{rhoads03}
Rhoads, J.~E., et al.\ 2003, AJ, 125, 1006

\bibitem[\protect\citeauthoryear{Santos et al.}{2004}]{santos04a}
Santos, M.R., Ellis, R.S., Kneib, J.-P., Richard, J., Kuijken, K.,
2004, ApJ, 606, 683

\bibitem[\protect\citeauthoryear{Santos}{2004}]{santos04b} Santos, M.,
2004, MNRAS, 349, 1137

\bibitem[\protect\citeauthoryear{Smith et al.}{2005}]{smith05}
Smith. G.P., Kneib, J.-P., Smail, I., Mazzotta, P., Ebeling, H.,
Czoske, O., 2005, MNRAS, 359, 417

\bibitem[\protect\citeauthoryear{Stanway, Bunker \&\
McMahon}{2003}]{sbm03} Stanway, E.R., Bunker, A.J., McMahon, R.G.,
2003, MNRAS, 342, 439

\bibitem[\protect\citeauthoryear{Stark et al.}{2007}]{stark07} Stark,
D.P., Ellis, R.S., Richard, J., Kneib, J.-P., Smith, G.P., Santos,
M.R., 2007, ApJ, 663, 10.

\bibitem[\protect\citeauthoryear{Taniguchi et al.}{2005}]{taniguchi05}
Taniguchi, Y., et al.\ 2005, PASJ, 57, 165

\bibitem[\protect\citeauthoryear{Willis \& Courbin}{2005}]{paper1}
Willis, J.P., Courbin, F., 2006, MNRAS, 357, 1348

\bibitem[\protect\citeauthoryear{Yan et al.}{2006}]{yan06} Yan, H.,
Dickinson, M., Giavalisco, M., Stern, D., Eisenhardt, P.~R.~M., \&
Ferguson, H.~C.\ 2006, ApJ, 651, 24

\end{thebibliography}
\end{document}